\documentclass[onecollarge]{svjour3}       
\smartqed  
\usepackage{psfrag,graphicx,epsfig,color}
\journalname{Journal of Statistical Physics}

\usepackage{graphicx}
\usepackage{color}
\usepackage{epsfig}
\usepackage{subfigure}
\usepackage[dvipsnames]{xcolor}

\usepackage{epstopdf}

\usepackage{tikz}
\usetikzlibrary{arrows,intersections}
\usepackage{pgfplots}
\usepackage[all]{xy}

\begin{document}

\title{document}
\title{Quantifying the lucky droplet model for rainfall}

\author{Michael Wilkinson}

\institute{Michael Wilkinson \at
             School of Mathematics and Statistics,\\
             The Open University,\\ 
             Walton Hall,\\ 
             Milton Keynes, MK7 6AA,\\
             England \\
             \email{m.wilkinson@open.ac.uk}}  

\date{Received: date / Accepted: date}

\maketitle

\begin{abstract}
It is difficult to explain rainfall from ice-free clouds, because the timescale 
for the onset of rain showers is shorter than the mean time for collisions 
between microscopic water droplets. It has been suggested that raindrops 
are produced from very rare \lq lucky' droplets, which undergo a large number of 
collisions on a timescale which is short compared to the mean time for a the first 
collision. This work uses large deviation theory to develop estimates 
for the timescale for the onset of a rain shower, as a function of the 
collision rate coefficients. 

The growth history of the fast-growing droplets which do become raindrops
is discussed. It is shown that their first few collisions are always approximately 
equally spaced in time, regardless of how the mean time for typical droplets varies as a 
function of the number of collisions. 
\end{abstract} 
\maketitle

\section{Introduction}
\label{sec: 1}

There are some fundamental challenges in understanding rainfall from clouds \cite{Mas71,Pru+97}.
When ice is present, the Bergeron mechanism \cite{Mas71} shows how condensation of water vapour 
onto ice crystals can  trigger precipitation, but when no ice is present rainfall relies 
on collision and coalescence of microscopic water droplets. It has proven 
to be difficult to understand how the rate of these collisions can be sufficient 
to explain rainfall \cite{Sha03}. 

The collisions occur mainly as a result of microscopic 
water droplets with different radius falling at different terminal velocities. 
An important aspect of the growth of raindrops is that, because larger droplets fall 
more rapidly and have a larger cross-section, the collisions happen more frequently 
as the droplet grows. If $\tau_n$ is the mean time between the first $n-1$ and the first $n$ 
collisions, it is expected that $\tau_n$ decreases rapidly as a function of $n$. According to a model 
which is discussed in detail below, the sum 
\begin{equation}
\label{eq: 1.1}
\langle T\rangle =\sum_{n=1}^{\cal N} \tau_n
\end{equation} 
which is the mean time for ${\cal N}$ collisions, approaches a finite limit as ${\cal N}\to \infty$. 
The typical radius of the microscopic droplets in clouds (which form on condensation nuclei such as 
salt particles from evaporated sea spray) is approximately $10\,\mu{\rm m}$, and 
the liquid water content of a dense cloud is 
approximately $1\,{\rm gm}^{-3}$ \cite{Mas71,Pru+97}. Using these data, the 
rate of collision of droplets is found to be very small. The mean time for the first collision, $\tau_1$, 
depends upon the dispersion of the droplet sizes, but it is much less than one collision per hour. 
A raindrop may may have a radius as large as $1\,{\rm mm}$. 
Because most collisions of the growing droplet are with microscopic droplets, 
approximately $10^6$ collisions are required to create a raindrop. It is common experience 
that newly formed clouds can generate rain showers quite rapidly, on a timescale of 
less than half an hour. Given the slow rate of the first few collisions, and the very large number 
of collisions that are required, it hard to understand how rainfall can happen so quickly.

The Lifshitz-Slyozov analysis of Ostwald ripening \cite{Lif+61,Lif+81} does not offer a resolution 
of this problem because it acts too slowly \cite{Cle08,Wil14}. 
It has also been pointed out that turbulence in cumulus clouds 
will enhance the collision rate \cite{Saf+56}, but extensive studies, reviewed in \cite{Pum+16}, 
indicate that this effect is not sufficient, except possibly in the most unstable atmospheric conditions.  
The most satisfying solution to this problem is the \lq lucky droplet' concept, originally 
proposed by Telford \cite{Tel55} and Twomey \cite{Two66}. These authors pointed out 
that it is those microscopic droplets which undergo their
first few collisions exceptionally quickly will grow into raindrops, and absorb a large number 
of other microscopic droplets in the process. In fact, the fastest growing 
droplets could absorb all the remaining microscopic water droplets on their way to becoming 
raindrops, so that the occurrence of a rain shower is determined by the growth history 
of the (approximately) one-in-a-million fastest growing droplets. 

Subsequently, Kostinski and Shaw \cite{Kos+05} formulated a model for the time taken for 
droplets to grow by collision, and presented  some simulations indicating that 
it could yield a satisfactory solution of the problem. 
It is desirable to have a simple and maximally transparent expression for this 
\lq lucky droplet' principle. In \cite{Wil16}, it was proposed that large deviation theory \cite{Fre+84,Tou09}
is the relevant tool. When $T$ is small, the probability $P(T)$ that a microscopic droplet undergoes runaway 
growth to become a rain droplet within time $T$ is exponentially small: write
\begin{equation}
\label{eq: 1.2}
P(T)= \exp[-{\cal J}(T)]
\end{equation}
where ${\cal J}(T)$ is closely related to the 
\emph{entropy function} or \emph{rate function} in large deviation theory. 
If a rain droplet is the result of coalescence of ${\cal N}$ microscopic droplets, then the 
fraction of the microscopic droplets which are removed after time $T$ is greater than 
${\cal N}P(T)$, provided the number of microscopic droplets has not yet been 
significantly reduced by collisions. The onset of the rain shower is at a time $T^\ast$ 
when a significant fraction of the microscopic water droplets have been removed 
by coalescence into falling rain drops. It will be argued that this time $T^\ast$ is estimated by writing 
${\cal N}\exp[-{\cal J}(T^\ast)]\sim 1$. The time $T^\ast$ at which a rain shower occurs is therefore 
estimated by solving the equation 
\begin{equation}
\label{eq: 1.3}
{\cal J}(T^\ast)=\ln\,{\cal N}
\ .
\end{equation}
The function ${\cal J}(t)$ diverges as $t\to 0$. 
It is, therefore, expected that, in the limit as ${\cal N}\to \infty$, the ratios $T^\ast/\langle T\rangle$ 
and $T^\ast/\tau_1$ become small. This implies that, if ${\cal N}$ is sufficiently large, 
a rain shower is expected to occur after a time which is small compared to the mean time 
for the first collision. 

Equation (\ref{eq: 1.3}) offers a surprisingly simple criterion for the onset of rain showers. 
If the brief arguments presented above are considered more carefully, as is done in section 
\ref{sec: 2} below, it can be argued that solving a variant of equation (\ref{eq: 1.3}) gives an 
upper bound on the time taken to produce a rain shower, in the context of a model for a 
homogeneous atmosphere.

In order to make practical use of equation (\ref{eq: 1.3}), ${\cal J}(T)$ should be 
expressed as a functional of the set of mean collision times $\tau_n$. 
This is addressed in section \ref{sec: 3}. 
Both of the earlier works which quantify the lucky droplet model, \cite{Kos+05} and \cite{Wil16}, 
assumed that $\tau_n$ has a power-law dependence upon $n$, writing $\tau_n=\tau_1n^{-\gamma}$, 
but it is desirable to determine ${\cal J}(T)$ in a more general case.  
Because the terminal velocity of a small sphere is proportional to the square of its radius $a$, and the 
collision cross section also increases as $a^2$, the collision rate is proportional to $a^4$ for droplets 
which have undergone a large number of collisions, but which have a terminal velocity with small Reynolds 
number. The volume of a droplet is proportional to the number of collisions that it has undergone, 
implying that $a^3\sim n$, so that the exponent is $\gamma=4/3$. 
However, this choice might not give a good description of the time between the first few collisions ($\tilde n$, say), 
which may occur at a different rate \cite{Mas71,Pru+97} (for example, collisions of very small droplets 
may be suppressed by lubrication effects). 
Accordingly, the following calculations emphasise the more general case where 
\begin{equation}
\label{eq: 1.4}
\tau_n=\tau_1 n^{-\gamma} \left[1+Q(n/\tilde n)\right]
\end{equation}
where $Q(x)$ is a positive function which approaches zero when $x\gg 1$, and which may be divergent 
as $x\to 0$. The objective is to give useful approximations to ${\cal J}(T)$
which are applicable in this more general case. The numerical illustrations will use 
\begin{equation}
\label{eq: 1.5}
Q(x)=x^{-\delta}\exp(-x)
\end{equation}
with $\gamma=4/3$ and $\delta>0$ (which implies that the first few collisions occur more 
slowly than predicted by the $\tau_n\sim n^{-4/3}$ relation). It will be shown that 
the time taken to initiate a rain shower is surprisingly insensitive to the first few values of the 
mean collision times $\tau_n$.

It is interesting to consider the growth history of the fastest growing droplets. 
This is addressed in section \ref{sec: 4}. Surprisingly, it is found that their  
first few collisions occur at approximately equally spaced time intervals, regardless of the form of dependence 
of $\tau_n$ upon $n$. To avoid unhelpful complication of the notation, section \ref{sec: 3} considers the 
growth history of droplets where the collision times are a simple power-law, $\tau_n=n^{-\gamma}$, rather 
than the more general case described by (\ref{eq: 1.4}). 

The formulae developed in sections \ref{sec: 3} and \ref{sec: 4} are compared with numerical 
simulations. Section \ref{sec: 5} is a concluding summary.

\section{Modelling rain showers}
\label{sec: 2}

Assume that there is, initially, a uniform density of microscopic 
water droplets in a homogeneous atmosphere. These undergo collisions, 
such that the density of droplets $\rho(m,t)\delta m$ in a small interval of 
mass, $[m,m+\delta m]$, varies as a function of time. 
Collision of a droplet of mass in the small interval $[m_1,m_1+\delta m_1]$ with another 
droplet having mass in the small interval $[m_2,m_2+\delta m_2]$ occurs with rate 
$K(m_1,m_2)\rho(m_1,t)\rho(m_2,t)\delta m_1\delta m_2$, where $K(m_1,m_2)$ is termed the collision kernel. 
The density of droplets $\rho(m,t)$ with mass $m$ after time $t$ is then specified by the Smoluchowski
equation \cite{Smo16,Mas71,Pru+97,Sha03,Ald99}: 
\begin{eqnarray}
\label{eq: 2.1}
\frac{\partial \rho(m,t)}{\partial t}&=&\frac{1}{2}\int_0^m{\rm d}m'\ K(m-m',m')\rho(m-m',t)\rho(m',t)
\nonumber \\
&&-\int_0^\infty {\rm d}m'\ K(m,m')\rho(m,t)\rho(m',t)
\ .
\end{eqnarray}
The collision kernel for small droplets falling under gravity has the form 
\begin{equation}
\label{eq: 2.2}
K(m_1,m_2)=A \varepsilon (m_1,m_2)(m_1^{1/3}+m_2^{1/3})^2|m_1^{2/3}-m_2^{2/3}|
\end{equation}
where $A$ is a constant and $\varepsilon(m_1,m_2)=\varepsilon(m_2,m_1)$ is a collision 
efficiency, which is assumed to approach unity after the first few collisions, but which may be 
small in the initial stages of droplet growth \cite{Mas71,Pru+97,Sha03}.

If the dispersion of the initial masses is small, the size of a droplet can be described by the number $n$ 
of droplets from which it has been formed by coalescence. It is this simplified picture which 
will be used in the following discussions, in which the initial un-collided droplets 
will be referred to as monomers, and the larger droplets as $n$-mers. 
The density of $n$-mers, $\tilde \rho(n,t)$, obeys a simplified Smoluchowski equation
\begin{eqnarray}
\label{eq: 2.3}
\frac{\partial \tilde \rho(n,t)}{\partial t}&=&\frac{1}{2}\sum_{m=1}^{n-1} \tilde K(n-m,m)\tilde \rho(n-m,t)\tilde \rho(m,t)
\nonumber \\
&&-\sum_{m=1}^\infty \tilde K(n,m)\tilde \rho(n,t)\tilde \rho(m,t)
\ .
\end{eqnarray}
Because the collision kernel, equation (\ref{eq: 2.2}), increases sufficiently rapidly as a function 
of the masses, the mean time to reach an infinite cluster size is finite: in the literature of the 
Smoluchowski equation, this runaway growth is termed a gelation transition \cite{Ald99}.
If the clusters are allowed to grow to infinite size, the Smoluchowski equation 
predicts that the gelation transition for the kernel (\ref{eq: 2.2}) occurs in zero time \cite{vDo87,Ald99,Ley03,Bal+11}. 
This is a consequence of the fact that the Smoluchowski equation is a mean-field 
approximation: in an infinite volume, there is one cluster that grows arbitrarily quickly, and 
this will absorb all of the other particles. 

In the application to modelling clouds, it will be assumed that the droplets 
grow to a finite size, achieved after ${\cal N}\approx 10^6$ microscopic droplets have coalesced.
The size limit could be a consequence of the finite depth of the cloud, or because raindrops
fragment due to aerodynamic forces above a certain size. 

If this upper cutoff is imposed, it is possible in principle solve the Smoluchowski equation 
to calculate the probability $P(T)$ that a given droplet has grown by coalescence to size 
${\cal N}$ after time $T$. Because the raindrop has absorbed ${\cal N}$ microscopic 
droplets in this process, if ${\cal N}P(T)=\mu$, with $0<\mu<1$, then at time $T$ a fraction 
of the liquid water content of the cloud greater than or equal to $\mu$  has become macroscopic 
rain droplets. It is, therefore, possible to estimate the time $T^\ast$ for the onset of a rain shower 
by solving the equation $P(T^\ast)=\mu /{\cal N}$. Under typical circumstances a rain shower 
converts a few percent of the liquid water content into rain, so taking $\mu >0.1$ (say) would 
give a reasonable upper bound on the time taken to produce a rain shower.

In the context of this approach to estimating $T^\ast$, the fact that the Smoluchowski 
equation is a mean-field equation is not a deficiency, because the period between 
collisions is sufficiently large that the system is \lq well-mixed'. However, it is extremely 
difficult to determine useful analytical expressions for solutions of the Smoluchowski equation, 
and even if these were available, it would be difficult to use them to determine $P(T)$.

An alternative approach is to consider the statistics of the time $T$ directly. A cluster 
of droplets grows to size $M$ by a sequence of collision events. The time between 
collision $j$ and the next collision will be denoted by $t_j$, and the time taken to 
undergo $M$ collisions is denoted by $T_M$.  The collision with index $j$ is with a particle 
of size $n_j$. The collisions are sufficiently infrequent that they may be regarded 
as independent events. Size ${\cal N}$ is reached at a time $T$ when the the following 
equations are satisfied: 
\begin{eqnarray}
\label{eq: 2.4}
T&=&\sum_{j=1}^M t_j
\nonumber \\
{\cal N}&\ge &\sum_{j=1}^M n_j
\ .
\end{eqnarray}
The size $n_j$ of the droplet which is absorbed at the $j^{\rm th}$ collision, and 
the time $t_j$ for this collision, are random variables. Their statistics are determined 
by proposing different random values of $t_j$ for each possible $n_j$. 
These will be denoted by $t^\dagger(n_j)$, 
and $t_j$ is determined by picking the minimum of these $t^\dagger(n_j)$ as the time until the next collision.  
The $t^\dagger(n_j)$ are all independent variables with a Poisson distribution, with PDF 
$p(t^\dagger)=\exp(-t^\dagger/\langle t^\dagger \rangle)/\langle t^\dagger \rangle$. If the collision with 
index $j$ occurs at time $T_j$ when the cluster size is $M$, then the mean values are
\begin{equation}
\label{eq: 2.5}
\langle t^\dagger(n_j) \rangle=\frac{1}{\tilde K(M,n_j)\tilde \rho(n_j,T_j)}
\ .
\end{equation}
Determining $T$ from equations (\ref{eq: 2.4}), (\ref{eq: 2.5}) is still a very complex task, 
and a further simplification will be applied. If only droplet growth by collisions with monomers 
are included, then the growth to size ${\cal N}$ will be slower, giving an upper bound of $T$ 
of the form 
\begin{eqnarray}
\label{eq: 2.6}
\hat T=\sum_{j=1}^{\cal N}t_j&&
\nonumber \\
p(t_j)=R_j\exp(-\tilde R_jt_j)
&\quad\quad\quad&
\tilde R_j=\tilde K(j,1)\tilde \rho(1,T_j)
\end{eqnarray}
 where $T_j$ is the time to reach size $j$. This quantity is still difficult 
 to calculate, because it is necessary to solve the Smoluchowski equation 
 to determine the time-dependent of the density of monomers, $\tilde \rho (1,t)$.
 An easier approach is to assume that the density of monomers does not decrease 
 to less than some fraction $\nu$ of its initial value before time $\hat T$, 
 that is  $\tilde \rho(1,t)\ge \nu \tilde \rho (1,0)$, with $0<\nu<1$, 
 for all $t\in [0,\hat T]$. The time $T$ is then bounded above by $\hat T/\nu$, where
 $\hat T$ is given by (\ref{eq: 2.4}), with $R_n=\tilde K(n,1)\tilde \rho(1,0)$. 
 
 To summarise, the time taken to produce a rain shower is less than the time $T^\ast$
 which solves 
 \begin{equation}
 \label{eq: 2.7}
 {\cal N}P(T^\ast \nu)=\mu
 \end{equation}
 where $\mu $ is the lower bound on the fraction of of the liquid water content which is 
 precipitated, and $1-\nu$ is the upper bound on the fraction of microscopic water 
 droplets which undergo collision. (Note that $\mu+\nu<1$). 
If $P(T)\sim \exp[-{\cal J}(T)]$, the time for onset of a shower is bounded by the solution 
of 
\begin{equation}
\label{eq: 2.8}
\ln {\cal N}-\ln \mu={\cal J}(T^\ast \nu)
\end{equation}
which is closely related to equation (\ref{eq: 1.3}). The function ${\cal J}(T)$ must diverge as $T\to 0$, 
so that when ${\cal N}$ is very large, the value of $T^\ast$ decreases when $\mu$ is decreased. 

The number ${\cal N}$ is very large, typically approximately $10^6$. In the limit as ${\cal N}\to \infty$, 
and for $\mu$, $\nu$ fixed numbers of order unity, the solution of (\ref{eq: 2.8}) for $T^\ast$ 
approaches zero. This indicates that $T^\ast$ can be much less than the mean time for the 
first collision, in which case $1-\nu$ will indeed be a small quantity. 

The problem of bounding the time taken to form a rain shower is, therefore, addressed  by calculating 
the function ${\cal J}(T)$. 

\section{Determining the entropy function}
\label{sec: 3}

\subsection{General approach}
\label{sec: 3.1}

The probability $P(T)$ that the sum (\ref{eq: 2.6}) is 
less than $T$ cannot be determined explicitly, but it can be related to the 
\emph{cumulant generating function}, $\lambda(k)$, defined by writing
\begin{equation}
\label{eq: 3.1}
\exp[-\lambda(k)]\equiv\langle \exp(-k T)\rangle= \int_0^\infty {\rm d}T\ p(T)\exp(-kT)
\end{equation}
where $p(T)=-{\rm d}P/{\rm d}T$ is the probability density function (PDF) of $T$.
The cumulant generating function $\lambda(k)$ can be determined exactly as a summation: noting 
that the $t_k$ are independent
\begin{equation}
\label{eq: 3.2}
\lambda(k)=\sum_{n=1}^{\cal N} \ln\left(1+k\tau_n\right)
\ .
\end{equation}
Note that $\exp[-\lambda(k)]$ is the Laplace transform of $p(T)$, which can be inverted 
by means of the Bromwich integral:
\begin{equation}
\label{eq: 3.3}
p(T)=\frac{1}{2\pi {\rm i}}\int_{\cal C} {\rm d}z\ \exp[zT-\lambda(z)]
\end{equation}
where ${\cal C}$ is a path in the complex plane with $z={\cal R}+{\rm i}\mu$, with $\mu$ a real number 
running from $-\infty$ to $\infty$, and ${\cal R}>-1/\tau_1$.  
This integral is estimated using the saddle point method: there is 
a saddle point at $z=k^\ast$ satisfying 
\begin{equation}
\label{eq: 3.4}
T=\sum_{n=1}^{\cal N} \frac{\tau_n}{1+k^\ast\tau_n}
\ .
\end{equation}
The PDF of $T$ is then approximated by 
\begin{equation}
\label{eq: 3.5}
p(T)\sim \frac{1}{\sqrt{2\pi |\lambda''(k^\ast)|}}\exp[-J(T)]
\end{equation}
where $J(T)$ is a Legendre transform of $\lambda(k)$:
\begin{equation}
\label{eq: 3.6}
J(T)=\lambda(k^\ast )-Tk^\ast
\ .
\end{equation}
Note that equation (\ref{eq: 3.5}) is very similar 
in form to (\ref{eq: 1.2}), implying that the Legendre transform function $J(T)$ is closely related to the large 
deviation entropy ${\cal J}(T)$ (defined by (\ref{eq: 1.2})) that is required. 
Noting that $P(T)=\exp[-{\cal J}(T)]$ is the integral of $p(t)$ 
from $0$ to $T$, and that within the region of integration $\exp[-J(T-\tau)]$ has its maximum at 
$\tau=0$, applying Watson's lemma to make an asymptotic estimate yields
\begin{eqnarray}
\label{eq: 3.7}
P(T)&=&\int_0^T {\rm d}t\ p(t)
\nonumber \\
&\sim&\int_0^T{\rm d}\tau\ \frac{1}{\sqrt{2\pi |\lambda''(k^\ast(T))|}}\exp[-J(T-\tau)]
\nonumber \\
&\sim &\frac{1}{\sqrt{2\pi|\lambda''(k^\ast(T))|}|J'(T)|}\exp[-J(T)]
\ .
\end{eqnarray}
By computing the function $J(T)$ equation (\ref{eq: 3.7}) can be used to determine ${\cal J}(T)$.

\subsection{Asymptotic approximation for the entropy function}
\label{sec: 3.2}

In order to write down explicit approximations for ${\cal J}(T)$, it is necessary to approximate  
the cumulant generating function by means of analytic functions. This can be achieved 
by approximating the summation of (\ref{eq: 3.2}) by an integral. The case in which 
$\tau_n=n^{-\gamma}$ is a foundation for the other examples. In \cite{Wil16}, the cumulant generating 
function for $\tau_n=n^{-\gamma}$ was obtained in the limit of as $k\to \infty$ and 
${\cal N}\to \infty$: the result is
\begin{equation}
\label{eq: 3.8}
\lambda(k)\sim \gamma A(\gamma) k^{1/\gamma} 
-\frac{1}{2}\ln k-\gamma {\cal C}+O(1/k)
\end{equation}
where 
\begin{equation}
\label{eq: 3.9}
A(\gamma)=\frac{1}{\gamma}\int_0^\infty {\rm d}x\ \frac{x^{-(\gamma-1)/\gamma}}{1+x}
\end{equation}
and where the remainder term ${\cal C}=0.918966\ldots$ can be expressed in terms of an infinite sum. 
The coefficient $A(\gamma)$ can be expressed in terms of the Euler beta function. 
For the important case where $\gamma=4/3$,  $A(4/3)=3\pi/(2\sqrt{2})$.

In this work equation (\ref{eq: 3.8}) will be extended to account for ${\cal N}$ being finite. Also, the 
formula will be extended to the case where the first few collisions 
occur with rates that do not conform to a power law, and which are given by  
equations (\ref{eq: 1.4}) and (\ref{eq: 1.5}). 

The correction due to finite ${\cal N}$ is easily obtained. Equation (\ref{eq: 3.8}) 
was obtained by taking the upper limit of the summation in (\ref{eq: 3.2}) to infinity. 
Subtracting the sum from ${\cal N}$ to $\infty$ from (\ref{eq: 3.8}), assuming 
that $k{\cal N}^{-\gamma}\ll 1$, and approximating the resulting summation 
by an integral, the correction  $\Delta \lambda(k)$ for finite ${\cal N}$ is
\begin{eqnarray}
\label{eq: 3.10}
\Delta \lambda (k)&=&\sum_{n={\cal N}+1}^\infty \ln\left[1+k\tau_n\right]
\sim \sum_{n={\cal N}+1}^\infty \ln\left[1+kn^{-\gamma}\right]
\nonumber \\
&\sim&k\sum_{n={\cal N}+1}^\infty n^{-\gamma}\sim \int_{n={\cal N}+1}^\infty {\rm d}n\ n^{-\gamma}
\nonumber \\
&\sim&k \frac{{\cal N}^{-(\gamma -1)}}{\gamma-1}
\ .
\end{eqnarray}
This is to be subtracted from (\ref{eq: 3.8}). This correction is
most significant in the case where $\gamma-1$ is small. 
It is straightforward to modify this calculation to deal with cases 
where the final stages of droplet growth conforms to a different 
relation from $\tau_n\sim n^{-\gamma}$, provided the summations in (\ref{eq: 3.10})
remains convergent as $n\to \infty$.

Next consider the correction due to replacing $\tau_n=n^{-\gamma}$ with equations 
(\ref{eq: 1.4}) and (\ref{eq: 1.5}). The consequent change in the cumulant generating function 
$\lambda  (k)$ is
\begin{eqnarray}
\label{eq: 3.11}
\Delta \lambda(k)&= &\sum_{n=1}^\infty \ln\left[\frac{1+kn^{-\gamma}[1+Q(n/\tilde n)]}{1+kn^{-\gamma}}\right]
\nonumber \\
&=&\sum_{n=1}^\infty \ln \left[1+\frac{Q(n/\tilde n)}{1+n^\gamma/k}\right]
\nonumber \\
&\sim&\sum_{n=1}^\infty \ln\left[1+Q(n/\tilde n)\left(1-\frac{n^\gamma}{k}\right)\right]
\nonumber \\
&\sim&\sum_{n=1}^\infty \ln[1+Q(n/\tilde n)]
-\frac{1}{k}\sum_{n=1}^\infty Q(n/\tilde n)n^\gamma
\ . 
\end{eqnarray}
Combining these gives a more refined version of equation (\ref{eq: 3.8}):
\begin{equation}
\label{eq: 3.12}
\lambda(k)\sim \gamma A(\gamma) k^{1/\gamma} 
-\frac{1}{2}\ln k-\gamma {\cal C}-\frac{k{\cal N}^{-(\gamma-1)}}{\gamma-1}
+\Sigma_1-\frac{\Sigma_2}{k}
\end{equation}
where 
\begin{equation}
\label{eq: 3.13}
\Sigma_1=\sum_{n=1}^\infty \ln\left[1+Q(n/\tilde n)\right]
\ ,\ \ \ 
\Sigma_2=\sum_{n=1}^\infty Q(n/\tilde n)n^\gamma
\ .
\end{equation}
In order to obtain the PDF of $T$, it is necessary to determine the Legendre transform of 
(\ref{eq: 3.11}). This requires the solution of (\ref{eq: 3.4}) to obtain $k^\ast(T)$. Using (\ref{eq: 3.11}) 
to approximate the sum in (\ref{eq: 3.4}), $k^\ast $ satisfies 
\begin{equation}
\label{eq: 3.14}
\tilde T\equiv T+\frac{{\cal N}^{-(\gamma-1)}}{\gamma -1}
=Ak^{-(\gamma-1)/\gamma}-\frac{1}{2k}+\frac{\Sigma_2}{k^2}
\ .
\end{equation}
The solution for small $T$ is
\begin{equation}
\label{eq: 3.15}
k^\ast(T)=\left(\frac{\tilde T}{A}\right)^{-\gamma/(\gamma-1)}
\left[1-\frac{\gamma\tilde T^{1/(\gamma-1)}}{2(\gamma-1)A^{\gamma/(\gamma-1)}}+\ldots \right]
\end{equation}
where $\tilde T$ is defined by (\ref{eq: 3.14}).
This yields an asymptotic expression for the probability density $p(T)$:
\begin{equation}
\label{eq: 3.16}
p(T)\sim K
\,\tilde T^{\frac{-(3\gamma-1)}{2(\gamma-1)}}\exp\left(-(\gamma-1)b\tilde T^{-1/(\gamma-1)}\right)
\end{equation}
where
\begin{equation}
\label{eq: 3.17}
b=A^{\gamma/(\gamma-1)}
\ ,\ \ \ 
K=\sqrt{\frac{\gamma}{2\pi(\gamma-1)}}\exp(\gamma {\cal C}-\Sigma_1)\,b
\end{equation}
and using equation (\ref{eq: 3.7}), the probability of undergoing ${\cal N}$ collisions in time less than $T$ is 
\begin{equation}
\label{eq: 3.18}
P(T)\sim \frac{K}{b\sqrt{\tilde T}}
\exp\left(-(\gamma-1)b\tilde T^{-1/(\gamma-1)}\right)
\ .
\end{equation}
This is an asymptotic formula which is valid in the limit as $T\to 0$, where $P(T)$ approaches 
zero very rapidly as $T$ decreases. Note that, because ${\cal J}(t)\equiv -\ln[P(t)]$, equation (\ref{eq: 3.18}) 
gives an explicit approximation for the function ${\cal J}(t)$ in equation (\ref{eq: 1.3}). 

\subsection{Numerical investigations}
\label{sec: 3.3}
 
The final result of the calculation, equation (\ref{eq: 3.18}), depends upon 
a sequence of asymptotic approximations, which may, or may not, work well 
in practice. These were tested by numerical experiments.  

\begin{figure}[!tbp]
  \centering
  \begin{minipage}[b]{0.49\textwidth}
    \includegraphics[width=\textwidth]{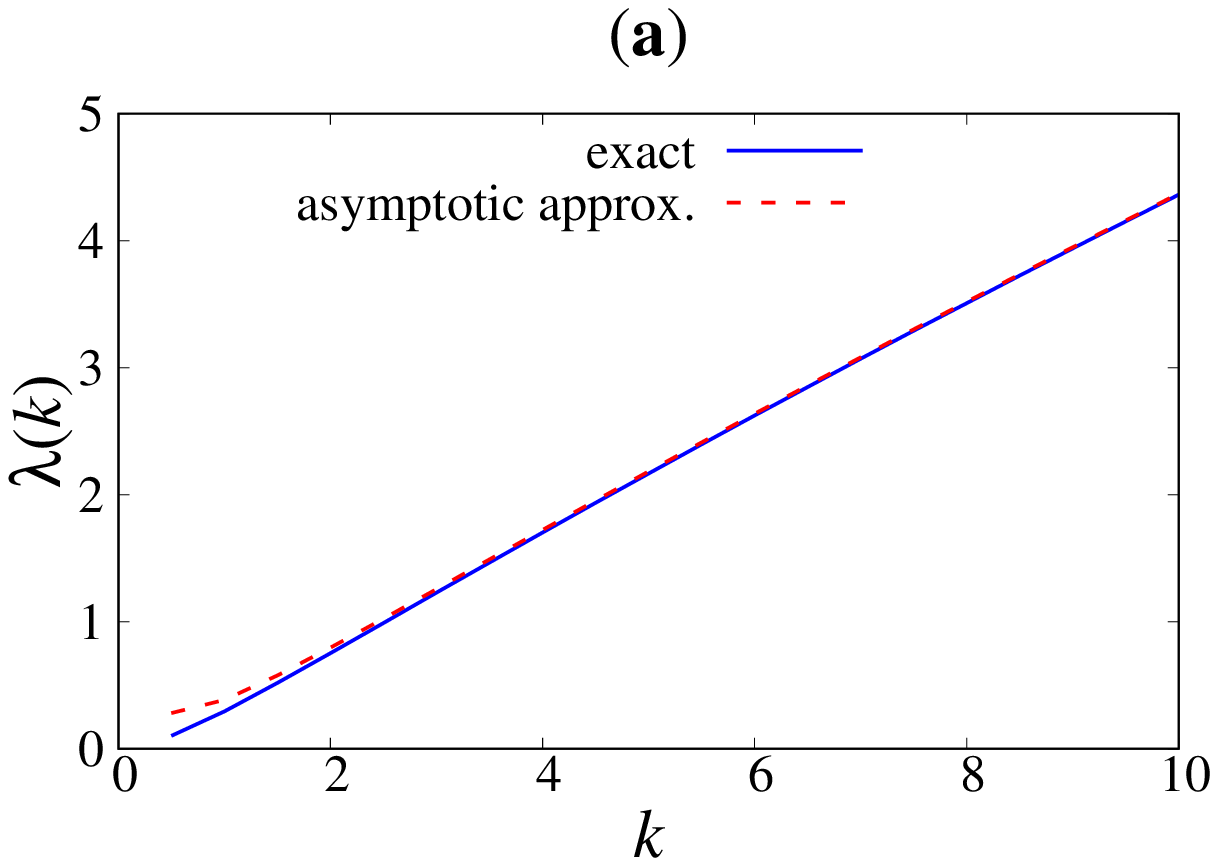}
  \end{minipage}
  \hfill
  \begin{minipage}[b]{0.49\textwidth}
    \includegraphics[width=\textwidth]{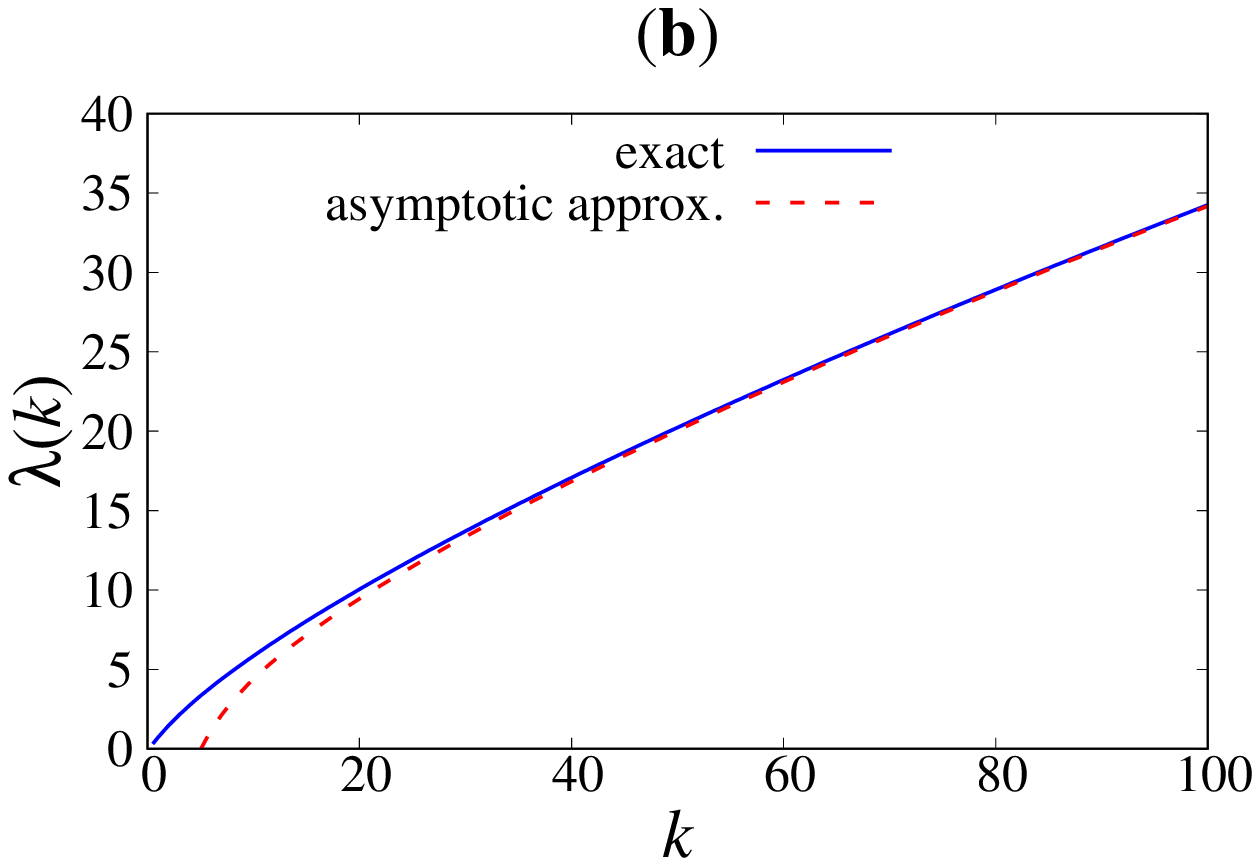}
  \end{minipage}
\caption{\label{fig: 1}
Cumulant generating function $\lambda(k)$, comparing exact value obtained by summation
(equation (\ref{eq: 3.2}) with asymptotic approximation, (\ref{eq: 3.12}). ({\bf a}): 
$\tau_n=n^{-4/3}$, ${\cal N}=10^4$. 
({\bf b}): $\tau_n=n^{-4/3}[1+(n/\tilde n)^{-2/3}\exp(-n/\tilde n)]$, $\tilde n=3$, ${\cal N}=10^4$.}  
\end{figure}

Figure \ref{fig: 1} compares the exact expression of $\lambda(k)$, equation (\ref{eq: 3.2}), 
with its asymptotic approximation, equation (\ref{eq: 3.12}). Figure \ref{fig: 1}({\bf a}) 
shows results for the case where $\tau_n=n^{-\gamma}$, with $\gamma=4/3$ and ${\cal N}=10^4$. 
Figure \ref{fig: 1}({\bf b}) displays results for $\tau_n$ given by (\ref{eq: 1.4}) and 
(\ref{eq: 1.5}), with $\gamma=4/3$, $\delta=2/3$, $\tilde n=3$ and 
${\cal N}=10^4$. There is excellent agreement as $k\to\infty$ in both cases, showing that the approximation
of $\lambda(k)$ by elementary analytic functions is very accurate. Correspondingly, the asymptotic approximation 
to $p(t)$ which was obtained from (\ref{eq: 3.12}) should be very accurate in the limit as $t\to 0$.

\begin{figure}[!tbp]
  \centering
  \begin{minipage}[b]{0.49\textwidth}
    \includegraphics[width=\textwidth]{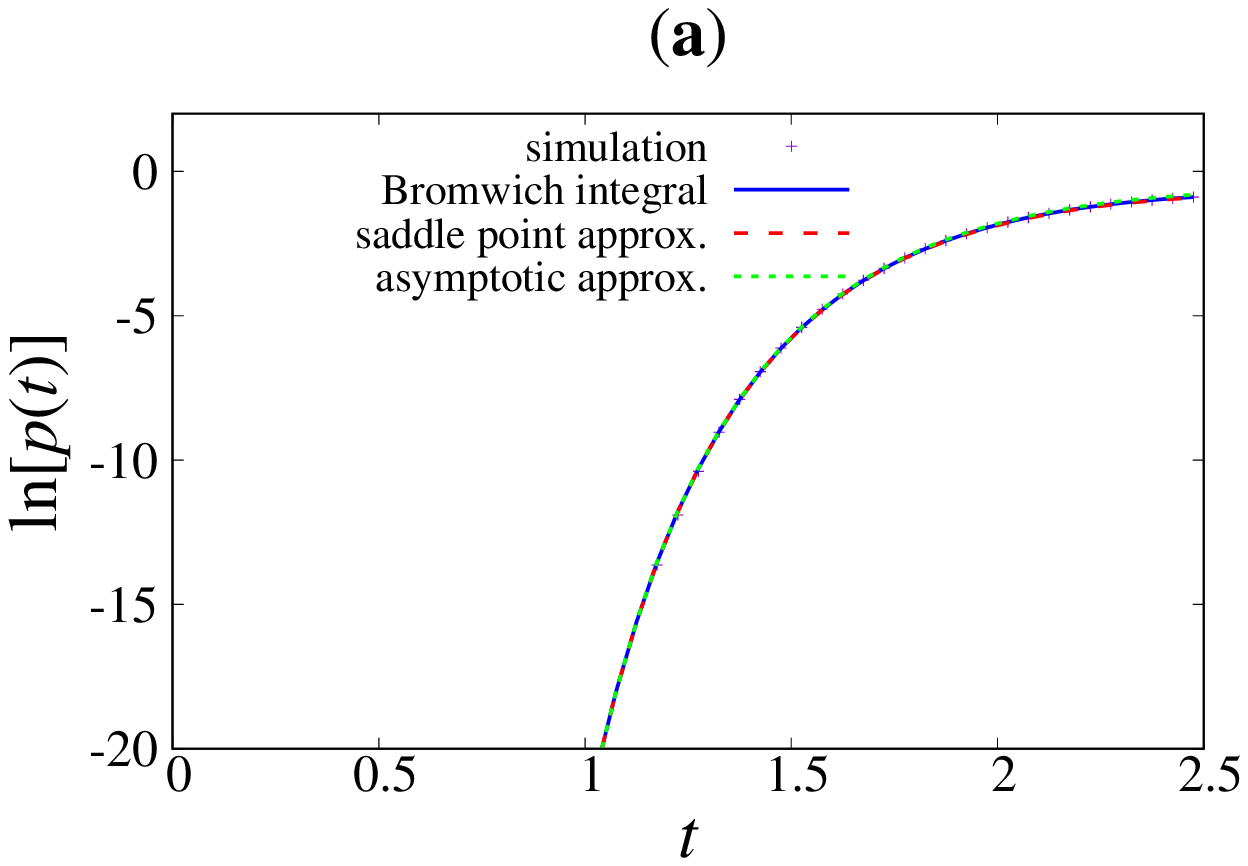}
  \end{minipage}
  \hfill
  \begin{minipage}[b]{0.49\textwidth}
    \includegraphics[width=\textwidth]{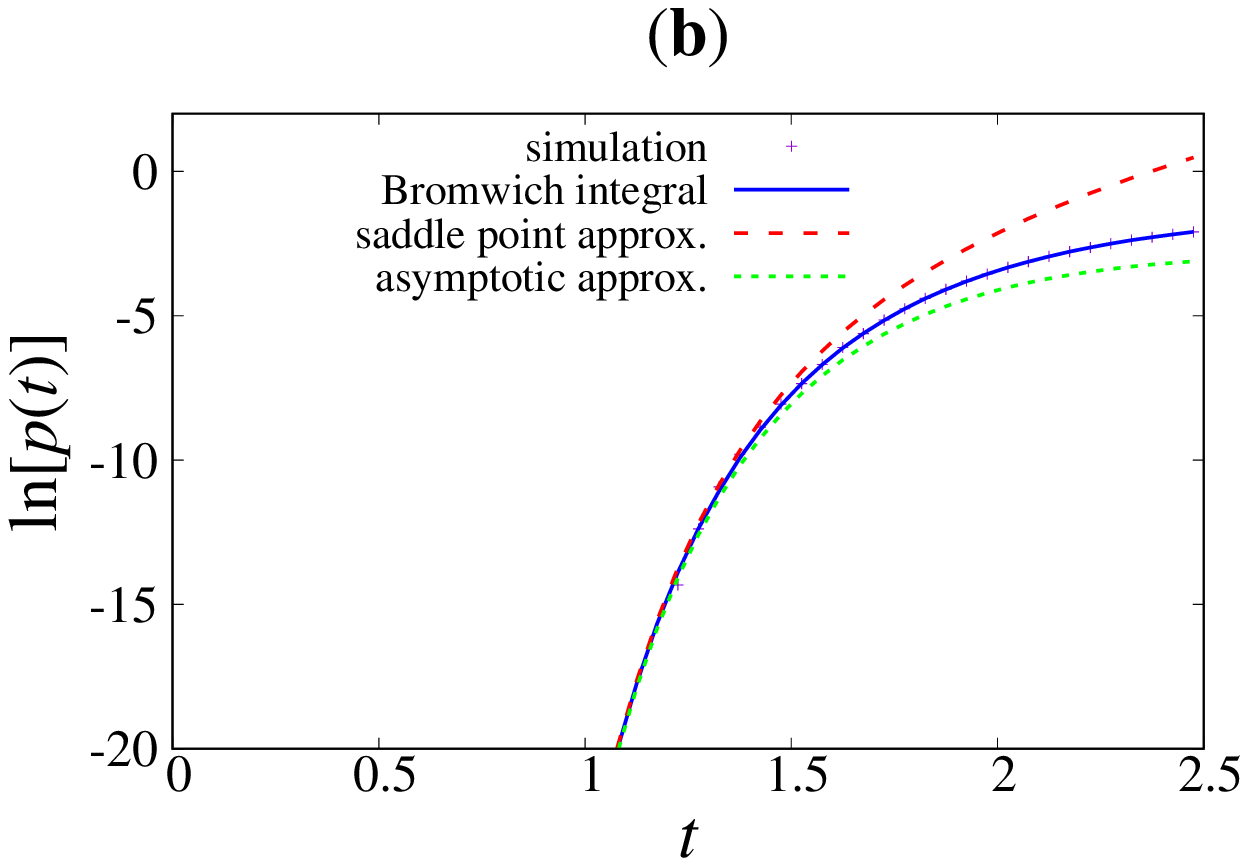}
  \end{minipage}
\caption{\label{fig: 2}
Comparing different approximations for the PDF $p(t)$ of the time taken for ${\cal N}$ collisions, 
according to the model (\ref{eq: 1.1}). Simulations and the Bromwich integral (\ref{eq: 3.2})
agree precisely. The saddle-point approximation (\ref{eq: 3.5}) and equation (\ref{eq: 3.16}) are 
in excellent agreement when $\tau_n=n^{-4/3}$, ${\cal N}=10^4$ (case ({\bf a})), and asymptotic 
to these other curves as $t\to 0$ when 
$\tau_n=n^{-4/3}[1+(n/\tilde n)^{-2/3}\exp(-n/\tilde n)]$, $\tilde n=3$, ${\cal N}=10^4$
(case ({\bf b})).
}  
\end{figure}

Figure \ref{fig: 2} compares different approaches to determining the PDF, $p(T)$, 
of the time taken to reach size ${\cal N}$. These are

\begin{enumerate}

\item Evaluation by simulation using ${\cal M}=10^8$ realisations of equation (\ref{eq: 1.1}).

\item Numerical evaluation of the Bromwich integral, equations (\ref{eq: 3.2}), (\ref{eq: 3.3}).

\item Saddle point approximation, equation (\ref{eq: 3.5}), with the cumulant generating function 
$\lambda(k)$ evaluated using the asymptotic formula (\ref{eq: 3.12}). The saddle-point equation 
(\ref{eq: 3.6}) is solved numerically to determine $k^\ast$.

\item Asymptotic approximation, using equation (\ref{eq: 3.16}).

\end{enumerate} 

The two panels show the same cases as figure \ref{fig: 1}. The first two methods 
should yield the same results, apart from fluctuations due to finite sample size of the 
simulation. The results of using the saddle-point approximation and equation (\ref{eq: 3.16}) are 
both in excellent agreement in the limit as $t\to 0$.

Figure \ref{fig: 3} compares estimates of $P(t)$ obtained by simulation (using $10^7$ realisations 
of (\ref{eq: 1.1})) with the result  
of the asymptotic formula, (\ref{eq: 3.18}), for a combination of parameter values which 
differs from figures \ref{fig: 1} and \ref{fig: 2}. Again, there is good agreement as $t\to 0$.
A notable feature of the figures \ref{fig: 2} and \ref{fig: 3} is that changing the rates of the 
initial collisions has very little effect. For example, comparing figure {\ref{fig: 3}({\bf b})
and figure \ref{fig: 2}({\bf a}), the mean time for the first collision, $\tau_1$, is increased 
by a factor of $1+4^{2/3}=3.519\ldots$, and the total number of collisions is increased to ${\cal N}=10^5$ 
from ${\cal N}=10^4$, but the intercept on the $t$ axis is only shifted by a small factor.

 \begin{figure}[!tbp]
  \centering
  \begin{minipage}[b]{0.49\textwidth}
    \includegraphics[width=\textwidth]{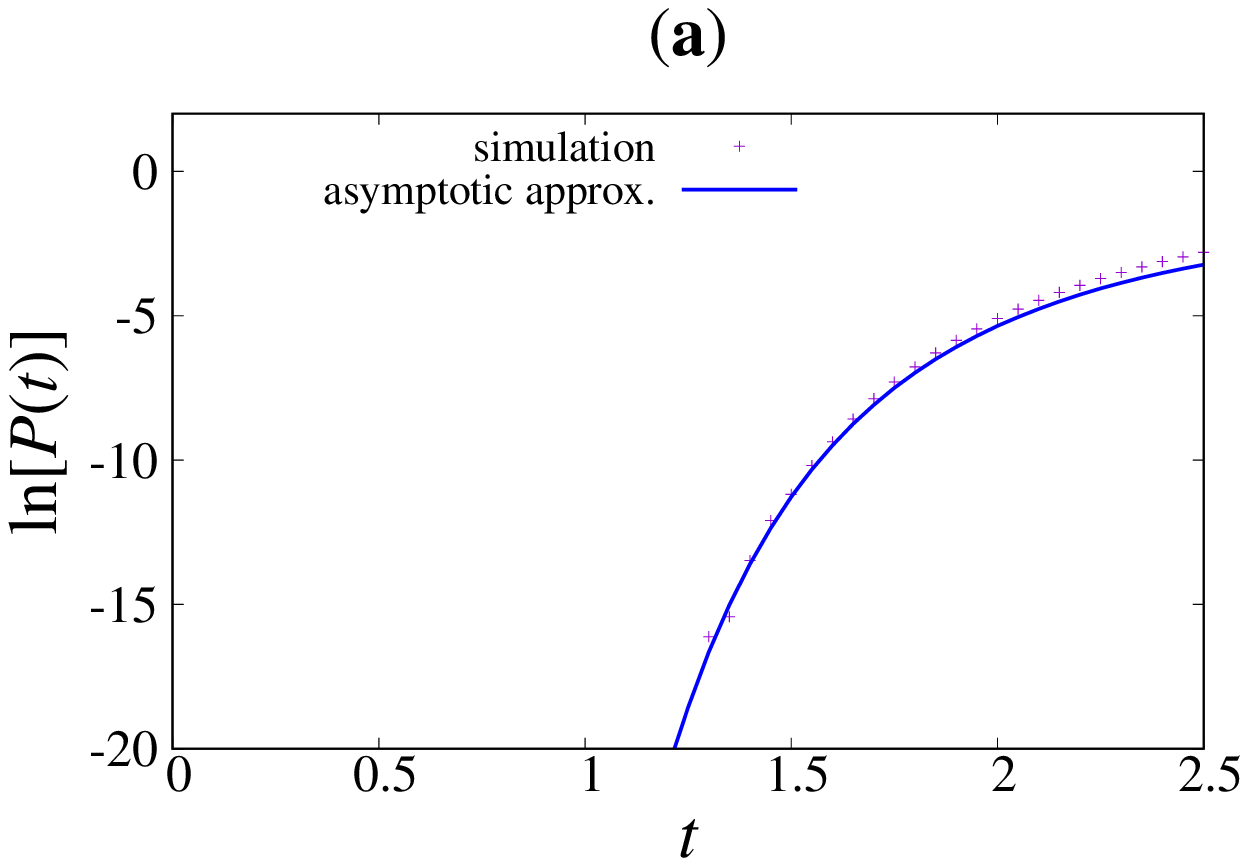}
  \end{minipage}
  \hfill
  \begin{minipage}[b]{0.49\textwidth}
    \includegraphics[width=\textwidth]{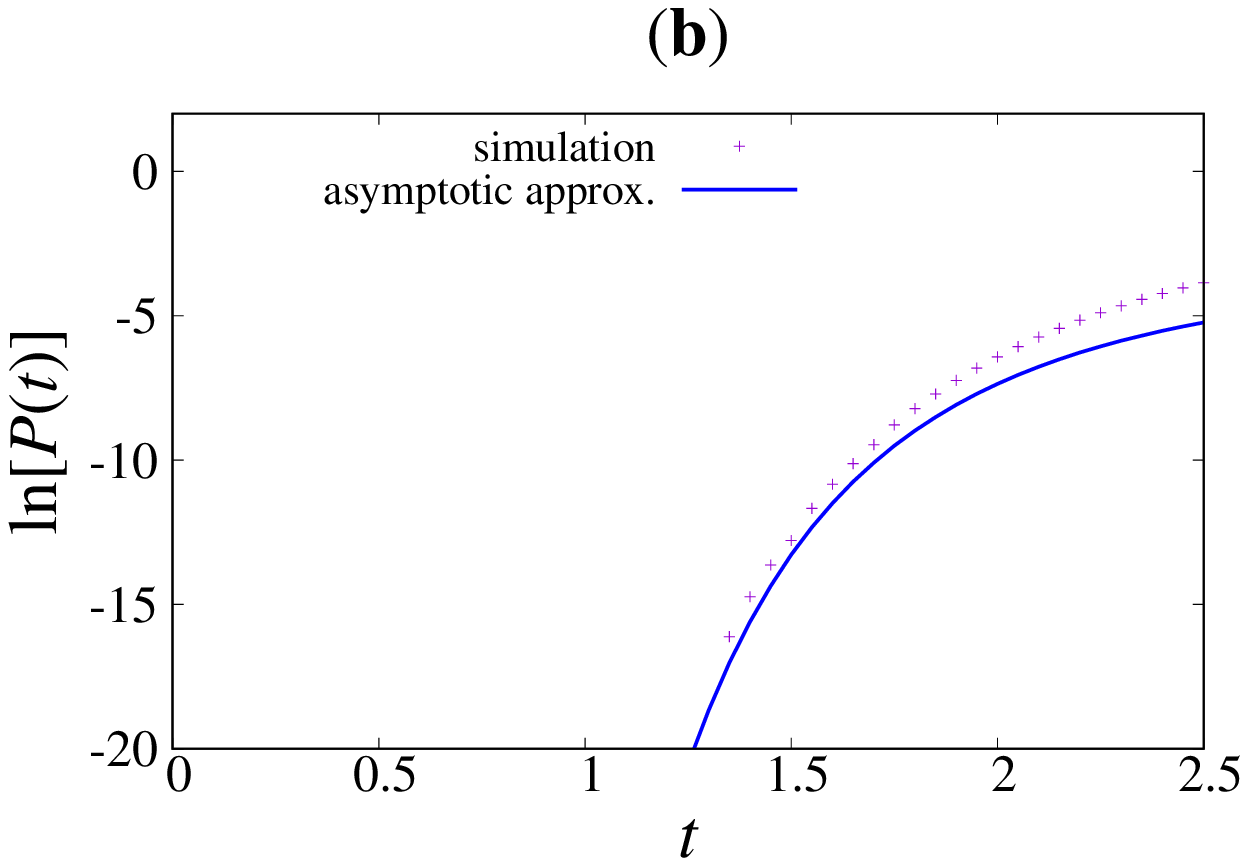}
  \end{minipage}
\caption{\label{fig: 3}
Comparing a simulation the probability $P(t)$ for a droplet to reach ${\cal N}$ collisions after time
$t$, with its asymptotic approximation, (\ref{eq: 3.18}): 
$\tau_n=n^{-4/3}[1+(n/\tilde n)^{-2/3}\exp(-n/\tilde n)]$, with ({\bf a}): $\tilde n=2$, ${\cal N}=10^5$,
({\bf b})  $\tilde n=4$, ${\cal N}=10^5$. 
}  
\end{figure}

\subsection{Implications for rainfall}
\label{sec: 3.4}
 
Section \ref{sec: 2} gave an equation, (\ref{eq: 2.8}), which gives an upper bound on the time 
$T^\ast$ to produce a rain shower, which removes a fraction greater than $\mu$ of the 
liquid water content of a cloud, subject to the constraint that the fraction of un-collided droplets 
is always greater than $\nu$ up until time $T^\ast$. Making use of this upper bound in the general 
case would require a complicated calculation, but there is a limiting case where (\ref{eq: 2.8}) can be 
replaced by a simpler relation, equation (\ref{eq: 1.3}). This is realised if the time $T^\ast$ is sufficiently small
that the fraction of un-collided droplets remains close to unity up to time $T^\ast$. At short times, the 
first collision of a droplet is almost always with another un-collided droplet, so that the fraction 
of un-collided droplets at time $t$ is approximated by  
\begin{equation}
\label{eq: 3.19}
\nu(t)=1-\frac{t}{\tau_1}+O(t^2)
\end{equation}
so that if the solution of (\ref{eq: 1.3}) satisfies $T^\ast/\tau_1\ll 1$, then the use of this 
simplified equation (\ref{eq: 1.3}) to estimate $T^\ast$ is justified.

Now that the large-deviation rate function is available from equation (\ref{eq: 3.18}), it is 
possible to estimate the ratio $T^\ast/\tau_1$ and establish when the use of the simplified 
equation (\ref{eq: 1.3}) is indeed justifiable (implying that a shower can occur in a timescale which is much 
shorter than the mean time for the first collision). From (\ref{eq: 3.18}), the rate function
may be written in the form
\begin{equation}
\label{eq: 3.20}
{\cal J}(T)=\alpha(\gamma)+\Sigma_1+\frac{1}{2}\ln \left(\frac{\tilde T}{T_0}\right)
+\left(\frac{\tilde T}{T_0}\right)^{-\frac{1}{\gamma-1}}
\end{equation}
where $\tilde T$ is defined by (\ref{eq: 3.14}), $\alpha(\gamma)$ is independent of $\tilde T$ and 
\begin{equation}
\label{eq: 3.21}
T_0(\gamma)=\left[A(\gamma)\right]^\gamma (\gamma-1)^{\gamma-1}
\ .
\end{equation}
In the important case where $\gamma=4/3$, $A(4/3)=3\pi/2\sqrt{2}$, $T_0(4/3)= 3.4508\ldots$
and $\alpha(4/3)=-0.38018\ldots$. 
Another case, for comparison, is $\gamma=2$, where $A(2)=\pi/2$, $T_0(2)=2.4674\ldots$
and $\alpha(2)=-0.81398\ldots$.

Now consider using these results to estimate $T^\ast$ by approximating the solution 
of (\ref{eq: 1.3}). An instructive approach is to approximate the rate function by 
${\cal J}(t)\approx \Sigma_1+(t/T_0)^{-1/(\gamma-1)}$, so that the solution 
to (\ref{eq: 1.3}) is approximated by
\begin{equation}
\label{eq: 3.22}
T^\ast\sim T_0\left[\ln ({\cal N}-\Sigma_1)\right]^{-(\gamma-1)}
\end{equation}
where ${\cal N}\approx 10^6$ is the number of microscopic droplets 
which coalesce to form a raindrop, and where $\Sigma_1$ is defined by equation (\ref{eq: 3.13}).
Note that this estimate for $T^\ast$ is rather insensitive to the rates of the first few collisions 
(which only enter through the quantity $\Sigma_1$).
This estimate for $T^\ast$ is to be compared with the timescale for the first collision, $\tau_1$, 
as specified by equation (\ref{eq: 1.4}), so that 
\begin{equation}
\label{eq: 3.23}
\frac{T^\ast}{\tau_1}\approx \frac{T_0}{1+Q(1/\tilde n)}\left[\ln{\cal N}-\Sigma_1\right]^{-(\gamma-1)} 
\ .
\end{equation}

In the case where the collision times follow a simple power-law, $\tau_n=n^{-\gamma}$, and $Q(x)=\Sigma_1=0$, 
the ratio $T^\ast/\tau_1$ is not a small number when $\gamma=4/3$ and ${\cal N}=10^6$: in that case equation 
(\ref{eq: 3.23}) gives $\frac{T^\ast}{\tau_1}\approx 1.438$, and a numerical solution of (\ref{eq: 1.3}) yields 
$\frac{T^\ast}{\tau_1}\approx 1.381$.  If, however, the first few collisions are much slower, this has a 
small effect on $T^\ast$, while $\tau_1$ can be greatly increased. For example, 
if ${\cal N}=10^6$ and $\tau_n=n^{-4/3}[1+(n/\tilde n)^{-2/3}\exp(-n/\tilde n)]$, with 
$\tilde n=5$, then $\Sigma_1=4.3671\ldots$ and equation (\ref{eq: 3.23}) gives $\frac{T^\ast}{\tau_1}\approx 0.428$, 
which is a fair approximation to the ratio obtained by a numerical solution of (\ref{eq: 1.3}), which yields 
$\frac{T^\ast}{\tau_1}\approx 0.460$. Because this ratio is less than unity, using the simplified condition, 
equation (\ref{eq: 1.3}) should give a fair approximation to the time taken to produce a rain 
shower. While it would be possible to make a more accurate estimate, the uncertainties 
arising from this estimate are less than those which are inherent in the model. 

\section{Distribution of time for passing through size $N$}
\label{sec: 4}

A cluster that undergoes runaway growth in time $T$ must pass 
through size $N$ (with $N<{\cal N}$), which is reached at some intermediate time $t_N<T$. 
The probability to undergo runaway growth in time $T$ may be written
\begin{equation}
\label{eq: 4.1}
P(T)=\int_0^T{\rm d}t_1\int_0^{t_1}{\rm d}t_2\ p^<_N(t_2) p^>_N(t_1-t_2)
\end{equation}
where $p^<_N(t)$ is the PDF for a cluster to grow from size $1$ to size $N$ 
in time $t$ and $p^>_N(t)$ is the PDF to grow from $N$ to ${\cal N}$ in time $t$.
Thus 
\begin{equation}
\label{eq: 4.2}
p(t)=\int_0^t{\rm d}t'\ p^<_N(t')p^>_N(t-t')
\end{equation}
is the probability density 
to have undergone runaway growth in time $t$. 
When $t$ is small, both of the probability densities 
$p^>_N$ and $p^<_N$ are very small, 
and can be addressed by large deviation theory: write
\begin{equation}
\label{eq: 4.4}
p^<_N(t)=\exp[-{\cal J}^<_N(t)]
\ ,\ \ \ 
p^>_N(t)=\exp[-{\cal J}^>_N(t)]
\ .
\end{equation}
In this limit it is, therefore, expected that the distribution $p^<_N(t') p^>_N(t-t')$ is very sharply 
peaked, with the maximum with respect to $t'$ at a point $t_N^\ast$. 
It is expected that those clusters which reach size ${\cal N}$ in a short time $T$ 
pass through size $N$ at a time which is close to $t_N^\ast$.
This time $t^\ast_N$ is well approximated by the position of the minimum of   
\begin{equation}
\label{eq: 4.5}
F_N(t)={\cal J}^<_N(t)+{\cal J}^>_N(T-t)
\ .
\end{equation}
 Now consider, in succession, the problem of determining the functions
 ${\cal J}^>_N$ and ${\cal J}^<$, before estimating $t^\ast_N$. 
 
 To avoid over-complicated notation, this calculation will be carried 
 out when the $\tau_n$ are a simple power-law: $\tau_n=n^{-\gamma}$.
 
\subsection{Runaway growth: $N$ to ${\cal N}$}
\label{sec: 4.1}

For power-law mean collision time $\tau_n=n^{-\gamma}$, 
the cumulant generating function for growth from $N$ to $\infty$ is 
\begin{equation}
\label{eq: 4.1.1}
\lambda^>_N(k)=\sum_{n=N}^{\cal N} \ln\left(1+kn^{-\gamma}\right)
\ .
\end{equation}
If $N=1$, this is well approximated by equation (\ref{eq: 3.8}).
Assuming that $kN^{-\gamma}\ll 1$, subtracting the sum from $1$ to $N$ gives
\begin{equation}
\label{eq: 4.1.2}
\lambda^>_N(k)\sim \gamma A(\gamma)k^{1/\gamma}-\left(N-\frac{1}{2}\right)\ln k-\gamma {\cal C}-\sigma_N
-k\frac{{\cal N}^{-(\gamma-1)}}{\gamma-1}
\end{equation}
with
\begin{equation}
\label{eq: 4.1.3}
\sigma_N=\sum_{n=1}^{N-1}\ln \tau_n
\ .
\end{equation}
Following the approach described in section \ref{sec: 3}, 
inverting the Laplace transform using a Bromwich integral, and approximating this 
using a saddle point (Laplace) approximation gives
\begin{equation}
\label{eq: 4.1.4}
p^>_N(t)=\frac{1}{\sqrt{2\pi |{\lambda^>_N}''(k^\ast)|}}\exp[-J^>_N(k^\ast)]
\ .
\end{equation}
The saddle point $k^\ast$ satisfies 
\begin{equation}
\label{eq: 4.1.5}
t=\frac{{\rm d}\lambda^>}{{\rm d}k}(k^\ast)
\ ,\ \ \ 
\frac{{\rm d}}{{\rm d}k}\lambda_N^>(k)=Ak^{(1-\gamma)/\gamma}-\frac{N-\frac{1}{2}}{k}
-\frac{{\cal N}^{-(\gamma-1)}}{\gamma-1}
\end{equation}
and $J^>_N(k)=\lambda^>_N(k)-k\lambda'^>_N(k)$.
The solution is, at leading order in $1/k$, 
\begin{equation}
\label{eq: 4.1.7}
k^\ast\sim b\tilde t^{-1/(\gamma -1)}\left(1-\frac{\gamma (N-\frac{1}{2})}{(\gamma -1)b}
\tilde t^{1/(\gamma-1)}\right)
\ ,\ \ \ \tilde t=t+\frac{{\cal N}^{-(\gamma-1)}}{\gamma-1}
\end{equation}
and $b$ is given by (\ref{eq: 3.17}).
Noting that $Ab^{1/\gamma}=b$, and expressing $J_N$ as a function 
of $t$ by writing $J_N(t)=J_N(k^\ast(t))$ gives
\begin{eqnarray}
\label{eq: 4.1.9}
J^>_N(\tilde t)&=&(\gamma-1)b \tilde t^{-1/(\gamma -1)}
+\frac{\gamma}{\gamma -1}\left(N-\frac{1}{2}\right)\ln \tilde t
\nonumber \\
&&-\left(N-\frac{1}{2}\right)\ln b-\gamma {\cal C}-\sigma_N
\end{eqnarray}
and hence
\begin{equation}
\label{eq: 4.1.10}
p^>_N(\tilde t)\sim \sqrt{\frac{\gamma}{2\pi(\gamma-1)}}b^N\exp[\gamma{\cal C}+\sigma_N)]
\tilde t^{-\frac{\gamma(2N+1)-1}{2(\gamma -1)}}
\exp\left(-(\gamma-1)b\tilde t^{-1/(\gamma-1)}\right)
\ .
\end{equation}
Upon setting $N=1$, this result is in agreement with (\ref{eq: 3.16}) when $\Sigma_1=0$. 
Figure \ref{fig: 4} compares equation (\ref{eq: 4.1.10}) with 
exact evaluation of $p^>_N(t)$ (simulation and evaluation of the Bromwich integral), and with 
the saddle point approximation, with $\gamma=4/3$ and ${\cal N}=10^4$, for $N=2$ 
({\bf a}) and $N=3$ ({\bf b}).

\begin{figure}[!tbp]
  \centering
  \begin{minipage}[b]{0.49\textwidth}
    \includegraphics[width=\textwidth]{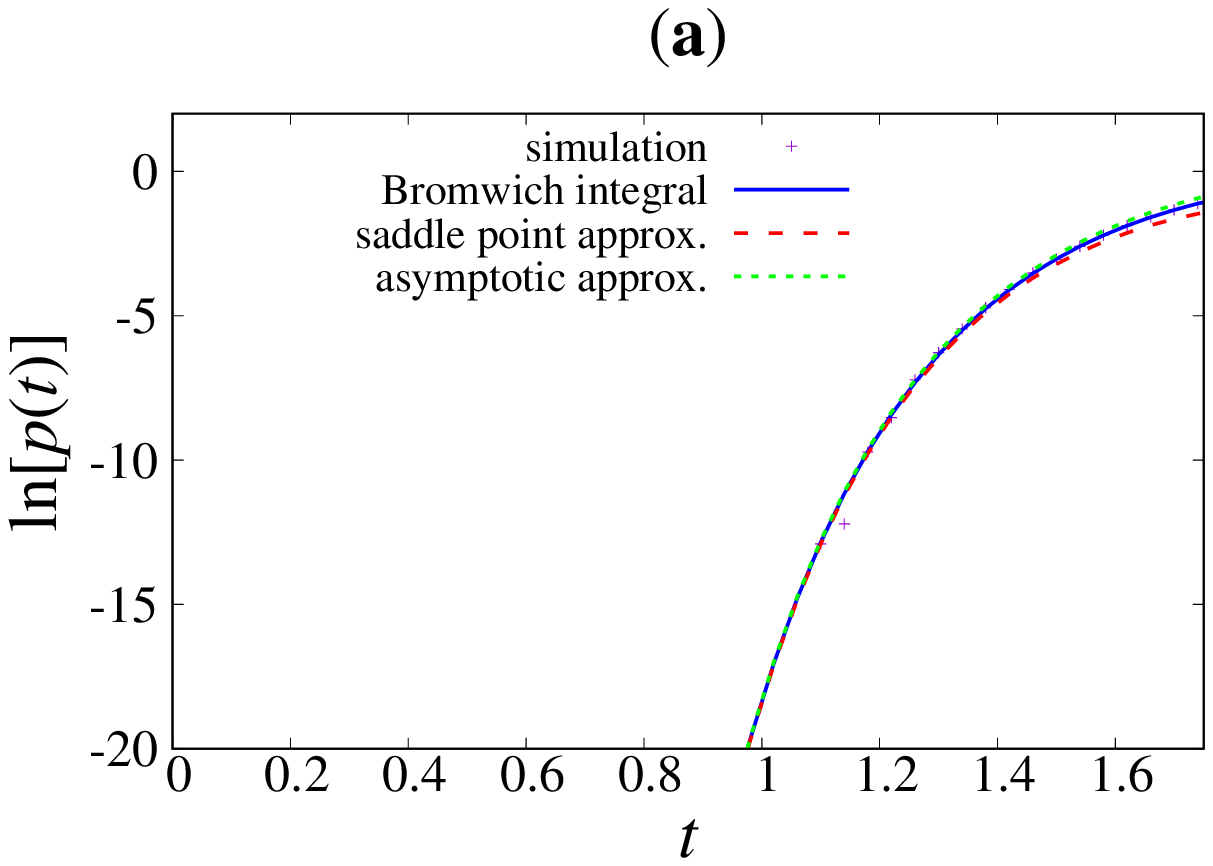}
  \end{minipage}
  \hfill
  \begin{minipage}[b]{0.49\textwidth}
    \includegraphics[width=\textwidth]{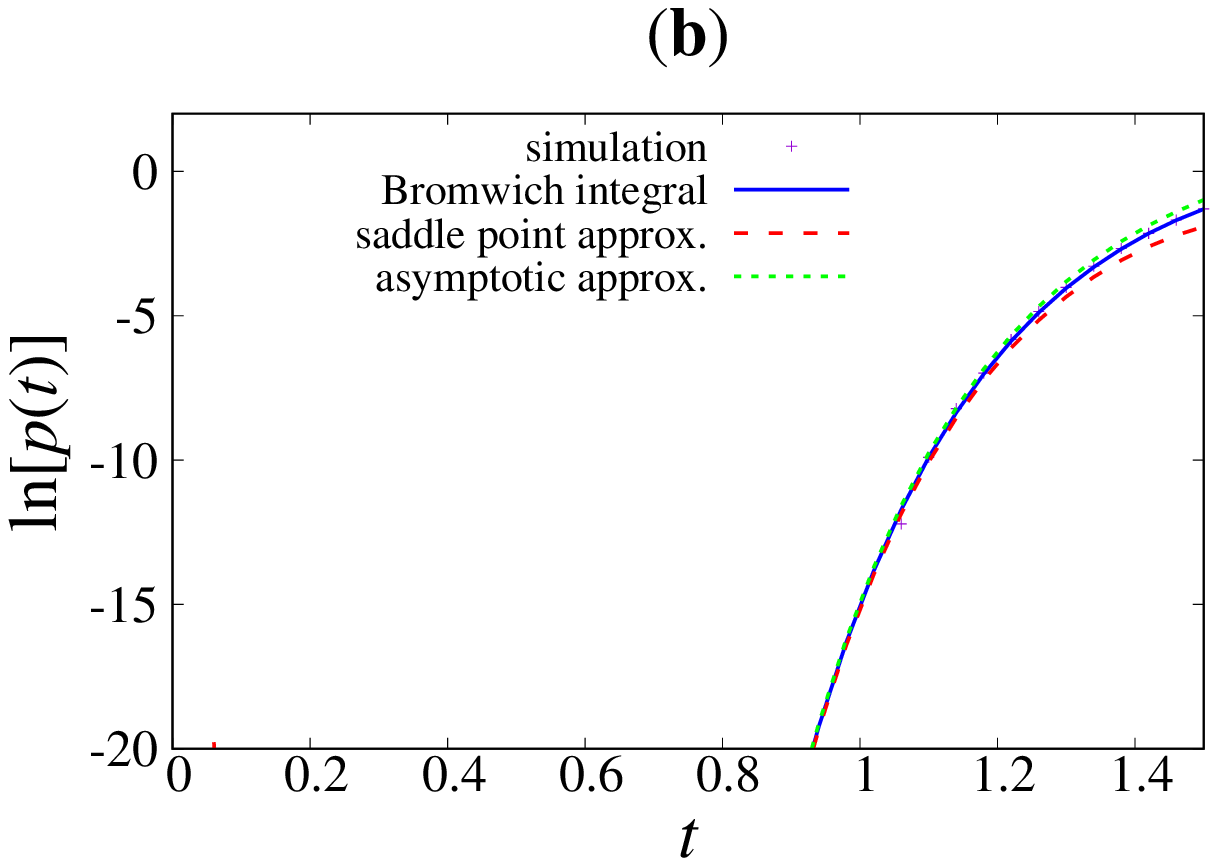}
  \end{minipage}
\caption{\label{fig: 4}
Comparing equation (\ref{eq: 4.1.10}) with simulation, the Bromwich integral (\ref{eq: 3.2}), and saddle-point 
approximation (\ref{eq: 3.5}). For both panels, $\tau_n=n^{-4/3}$, ${\cal N}=10^4$.
({\bf a}):  $N=2$, 
({\bf b}): $N=3$.
}  
\end{figure}

\subsection{Initial growth: $1$ to $N-1$}
\label{sec: 4.2}

Now estimate the probability density $p^<_N(t)$ to undergo the first $N-1$ collisions 
in time $t$, using the same approach. The cumulant generating function is  
\begin{equation}
\label{eq: 4.2.1}
\lambda^<_N(k)=\sum_{n=1}^{N-1} \ln\left(1+kn^{-\gamma}\right)
\ .
\end{equation}
To leading order in $k^{-1}$, this is
\begin{equation}
\label{eq: 4.2.2}
\lambda^<_N(k)\sim (N-1)\ln k -\sigma_N
\end{equation}
where $\sigma_N$ was defined in (\ref{eq: 4.1.3}).
The saddle point satisfies 
\begin{equation}
\label{eq: 4.2.3}
t=\frac{{\rm d}\lambda^<_N}{{\rm d}k}(k^\ast)
\end{equation}
so that 
\begin{equation}
\label{eq: 4.2.4}
k^\ast=\frac{N-1}{t}
\end{equation}
and hence $\frac{{\rm d}^2\lambda^<_N}{{\rm d}k^2}(k^\ast)=-t^2/(N-1)$. As a function of $t$, at 
leading order,  
\begin{equation}
\label{eq: 4.2.5}
J^<_N(t)\sim -(N-1)\ln t+(N-1)\left[\ln (N-1)-1\right]+\sigma_N
\end{equation}
so that 
\begin{equation}
\label{eq: 4.2.6}
p^<_N(t)\sim \sqrt{\frac{N-1}{2\pi}}\exp\left[-\sigma_N-(N-1)\left(\ln (N-1)-1\right)\right]t^{N-2} 
\ .
\end{equation}
Figure \ref{fig: 5} compares equation (\ref{eq: 4.2.6}) with 
exact evaluation on $p^<_N(t)$ (simulation and evaluation of the Bromwich integral), and with 
the saddle point approximation, with $\gamma=4/3$ and ${\cal N}=10^4$, 
for $N=2$ ({\bf a}) and $N=3$ ({\bf b}).

\begin{figure}[!tbp]
  \centering
  \begin{minipage}[b]{0.49\textwidth}
    \includegraphics[width=\textwidth]{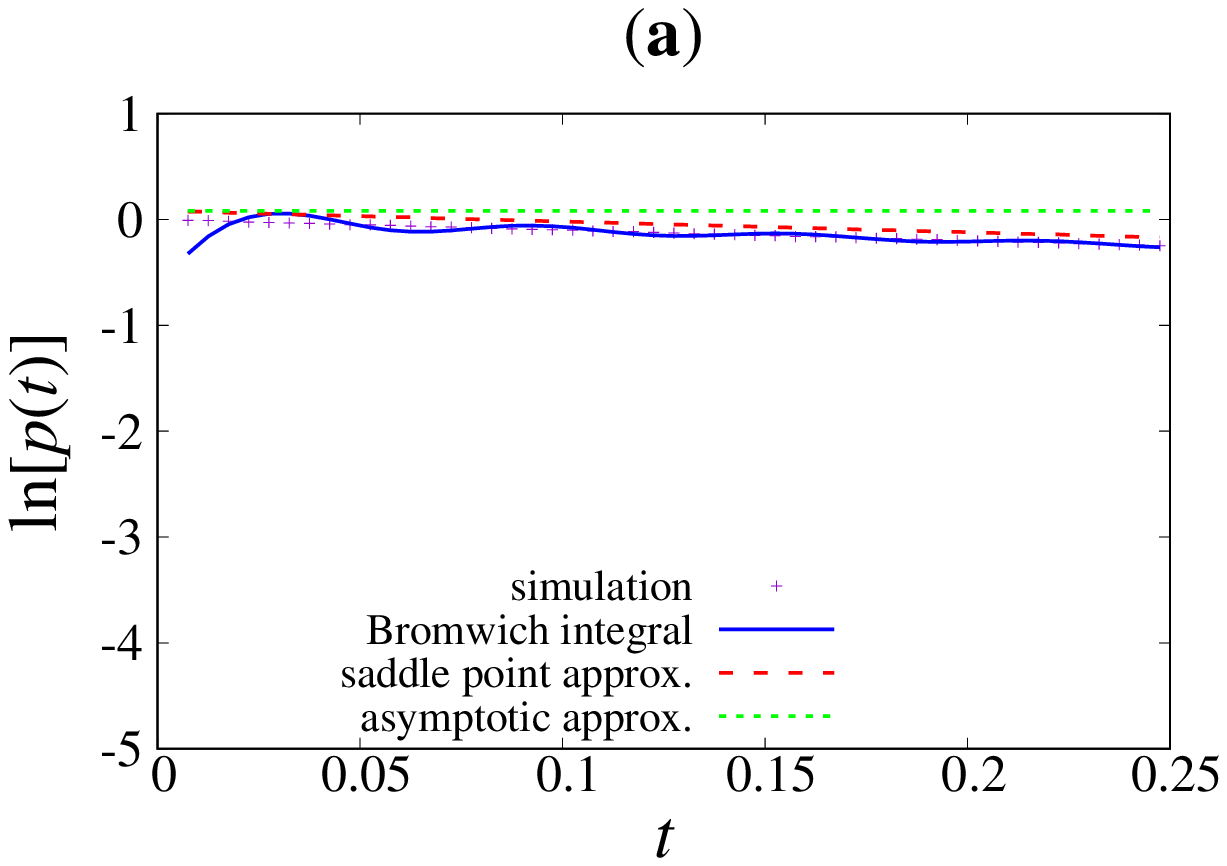}
  \end{minipage}
  \hfill
  \begin{minipage}[b]{0.49\textwidth}
    \includegraphics[width=\textwidth]{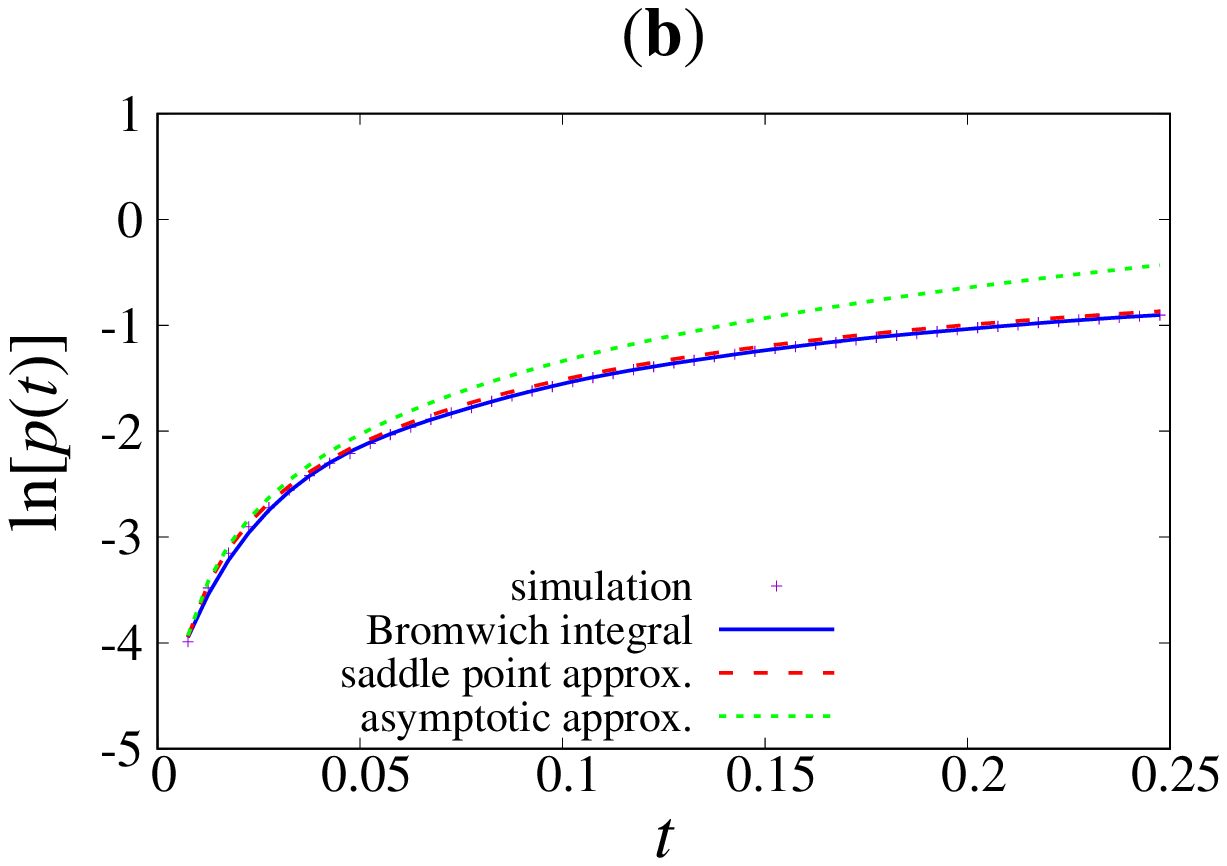}
  \end{minipage}
\caption{\label{fig: 5}
Comparing equation (\ref{eq: 4.2.6}) with simulation, the Bromwich integral (\ref{eq: 3.2}), and saddle-point 
approximation (\ref{eq: 3.5}). For both panels, $\tau_n=n^{-4/3}$, ${\cal N}=10^4$.
({\bf a}):  $N=2$, 
({\bf b}): $N=3$.
}  
\end{figure}

\subsection{Growth history of lucky drops}
\label{sec: 4.4}

The forms of the functions ${\cal J}^<_N$ and ${\cal J}^>_N$ can now be identified. 
Ignoring irrelevant constant terms, the function defined in (\ref{eq: 4.5}) is
\begin{eqnarray}
\label{eq: 4.4.1}
F_N(\tilde t)&=&-(N-2)\ln \left(\tilde t -\frac{{\cal N}^{-(\gamma-1)}}{\gamma-1}\right)
+b(\gamma-1)(T-\tilde t)^{-1/(\gamma-1)}
\nonumber \\
&&+
\frac{\gamma(2N+1)-1}{2(\gamma-1)}\ln(T-\tilde t)
\end{eqnarray}
where $\tilde t$ was defined in (\ref{eq: 4.1.7}).
Note that there are two terms which are functions of $T-\tilde t$, and that in the limit as $T\to 0$, 
the power-law term is dominant over the logarithmic term. Ignoring the term 
in $\ln (T-\tilde t)$, and differentiating the remaining terms with respect to $\tilde t$
\begin{equation}
\label{eq: 4.4.2}
F'_N(\tilde t)\sim b(T-\tilde t)^{-\gamma/(\gamma-1)}-\frac{(N-2)}{\tilde t-{\cal N}^{-(\gamma-1)}/(\gamma-1)}
\end{equation}
so that, if $t_N/T\ll 1$, the minimum of $F_N(t)$ is at 
\begin{equation}
\label{eq: 4.4.3}
t_N^\ast \sim \frac{N-2}{b}T^{\gamma/(\gamma-1)}
+\frac{{\cal N}^{-(\gamma-1)}}{\gamma-1}
\ .
\end{equation}
This implies that, for those \lq lucky' droplets that do undergo runaway
growth, the typical value $t^\ast_N$ of the time when they have undergone $N$ 
collisions is initially increasing linearly with $N$, regardless of the value of the exponent $\gamma$. 
This is a surprising result. It is, however, consistent with the observations made in section
\ref{sec: 3}, that the growth time of the fast-growing droplets is not strongly influenced 
by the first few values of the $\tau_n$.

The approximation (\ref{eq: 4.4.3}) breaks down when the predicted value
of $t^\ast_N$ is no longer small compared to $T$. This occurs after $N^\ast$ collisions, 
where $N^\ast \sim bT^{-1/(\gamma -1)}$.

The history of the fastest growing droplets was investigated numerically. 
The growth process represented by (\ref{eq: 1.1}) was simulated for 
${\cal M}$ realisations, and the random number seeds for the $K$ 
realisations where the sum (\ref{eq: 1.1}) is less than $T$ were recorded. 
These $K$ cases were run again, and their growth history $N(t)$ was recorded 
(where $N$ is the size of the cluster at time $t$).
Figure \ref{fig: 6} plots the mean size $\langle N(t)\rangle$ (and also the median size) of these $K$ clusters, 
which represent the $K/{\cal M}$ fraction which exhibits the fastest growth.
Simulations with $\tau_n=n^{-\gamma}$, for two different values of $\gamma$, both 
show a linear initial growth. 
The mean value $\langle N(t)\rangle$ is compared with the prediction
\begin{equation}
\label{eq: 4.4.5}
\langle N(t)\rangle\sim bT^{-\gamma/(\gamma-1)}t+{\rm const.}
\ .
\end{equation}
obtained from equation (\ref{eq: 4.4.3}), and good agreement is observed.

\begin{figure}[!tbp]
  \centering
  \begin{minipage}[b]{0.49\textwidth}
    \includegraphics[width=\textwidth]{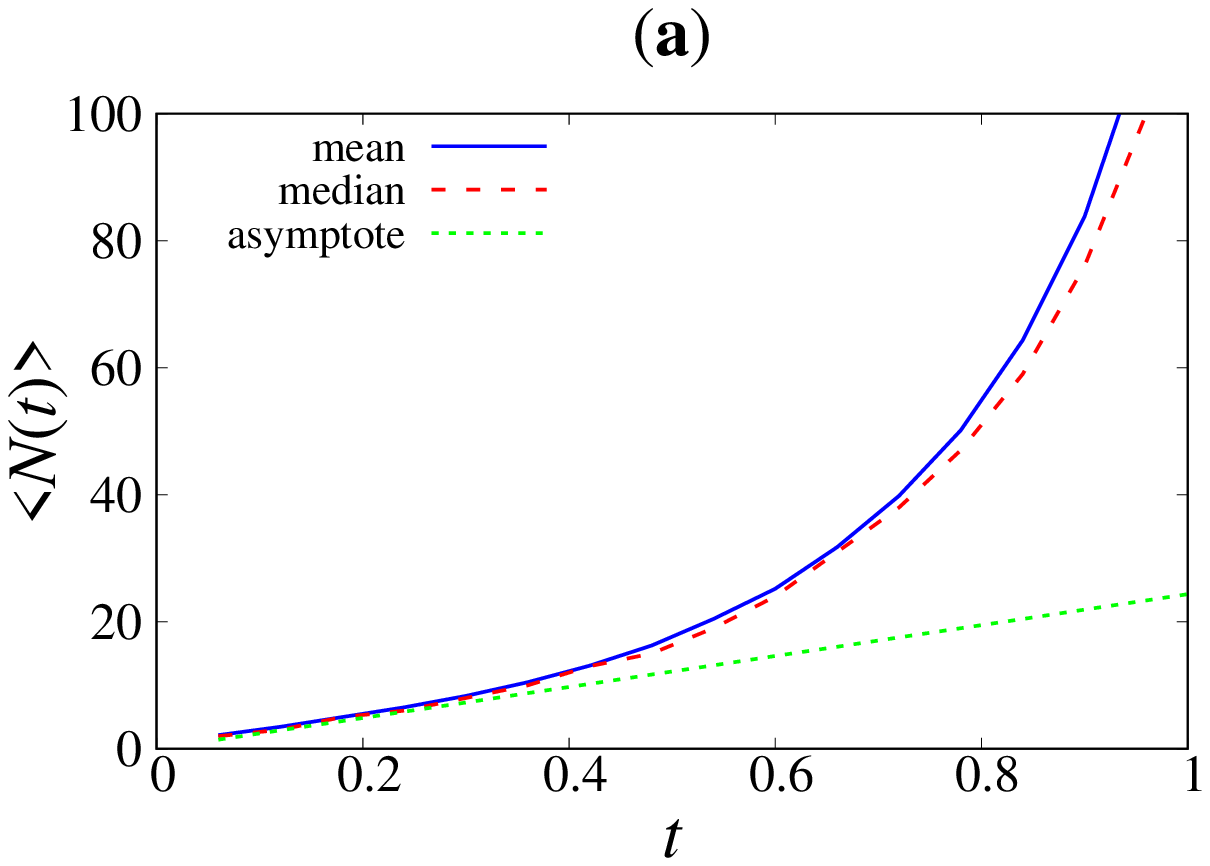}
  \end{minipage}
  \hfill
  \begin{minipage}[b]{0.49\textwidth}
    \includegraphics[width=\textwidth]{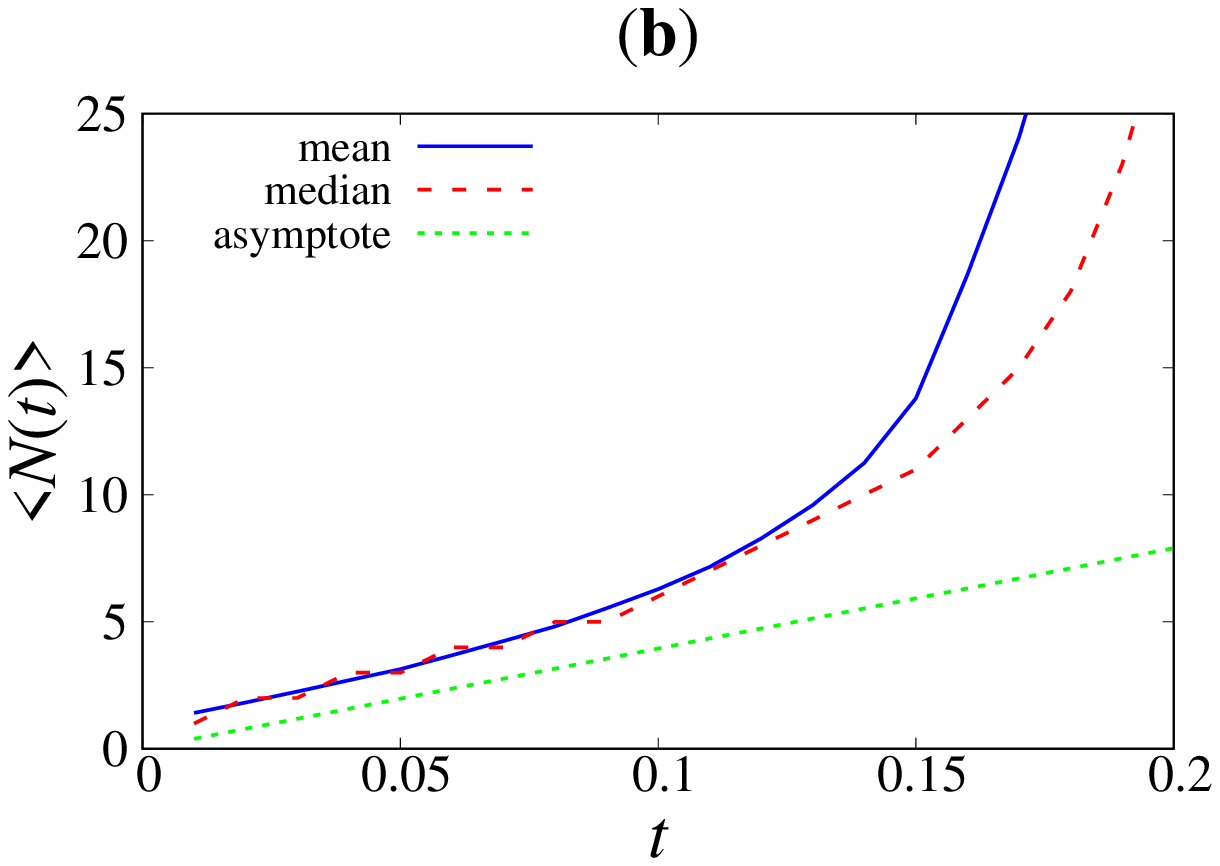}
  \end{minipage}
\caption{\label{fig: 6}
Plot of mean particle size $\langle N(t)\rangle$ for the fraction $K/{\cal M}\approx P(T)$ of 
fastest growing particles, which reach size ${\cal N}$ in time less than $T$. 
({\bf a}):  $\gamma=4/3$, $T=1.5$, $P(T)=1.884\ldots\times 10^{-4}$, ${\cal N}=10^4$. 
({\bf b}):  $\gamma=2$, $T=0.25$, $P(T)= 3.681\ldots\times 10^{-4}$, ${\cal N}=10^4$. 
In both cases the number of realisations of equation (\ref{eq: 1.1}) was ${\cal M}=10^6$.
}  
\end{figure}

\section{Conclusions}
\label{sec: 5}

The slow rate of collision between microscopic droplets in ice-free (\lq warm') clouds 
appears to impose a severe difficulty in explaining the rapid onset of rain showers.
However, this difficulty \emph{may} be overcome by recognising that, because a raindrop
is the result of coalescence of roughly a million microscopic droplets, a rain shower is 
created by the one-in-a-million fraction of droplets which have the fastest growth. 

Reference \cite{Wil16} proposed a surprisingly simple equation, (\ref{eq: 1.3}), for 
determining the time taken for the onset of a rain shower. This work has developed the 
arguments supporting equation (\ref{eq: 1.3}) in greater detail. It has also shown how the 
function ${\cal J}(T)$ can be approximated in cases where the mean collision times $\tau_n$ 
have a much more general dependence on $n$ than the simple power-law considered in \cite{Wil16}.
The probability $P(t)$ for undergoing ${\cal N}$ collisions in time $t$ is given by 
an explicit asymptotic expression, equation (\ref{eq: 3.18}), which is valid for a 
rather general form of the collision times $\tau_n$. This formula implies that 
the timescale for the onset of a rain shower can be short compared to the 
mean time for even the first of the million collisions.  

Equation (\ref{eq: 3.18}) allows a quantitative application to understanding rainfall from 
a homogeneous atmosphere. However, its principal significance for meteorology may be 
qualitative rather than quantitative, in that it implies that the apparent kinetic barrier to making 
rainfall from warm clouds is of little significance, even for a homogeneous atmosphere. 
In practice, the atmosphere may
contain a sufficiently large fraction of unusual nucleation centres, which nucleate atypically large 
microscopic droplets. It is these larger droplets which are likely to become raindrops.

While not readily experimentally observable, it is of interest to understand the growth history of those 
rare, fast-growing droplets that do become raindrops. Rather surprisingly, they are found to initially grow 
linearly as a function of time, irrespective of the $n$-dependence of the mean collision times, $\tau_n$.

The datasets generated during and/or analysed during the current study are available 
from the corresponding author on reasonable request.

\end{document}